\documentclass[traditabstract]{aa}

\usepackage{graphicx}
\usepackage[]{subfigure}
\usepackage{epsfig}
\usepackage{graphicx}
\usepackage{enumerate}
\usepackage{txfonts}
\usepackage{natbib}
\bibpunct{(}{)}{;}{a}{}{,} 

\newcommand{\nc}{\newcommand}
\nc{\mb}{\mathbf}
\nc{\LCDM}{$\Lambda$CDM}
\nc{\wcdm}{$w$CDM+$k$}
\nc{\mbf}{\mathbf}
\nc{\mnu}{$m_\nu$}
\nc{\snu}{$\nu_s$}
\begin{document}

\title{Cosmology with sterile neutrino masses from oscillation experiments}
\author{J. R. Kristiansen \inst{\ref{hioa}} \thanks{\email{joskri@hioa.no}}
\and \O. Elgar\o y \inst{\ref{astro}}    
\and C. Giunti \inst{\ref{giunti}}
\and M. Laveder \inst{\ref{laveder}}}
\institute{Oslo and Akershus University College of Applied Sciences, Faculty of Technology, Art and Design, Box 4 St. Olavs plass, N-0130 Oslo, NORWAY \label{hioa}
\and Institute of theoretical astrophysics, University of Oslo, Box 1029, N-0315 Oslo, NORWAY \label{astro}
\and Department of Physics, University of Torino and INFN, Via P. Giuria 1, I-10125 Torino, Italy \label{giunti}
\and Dipartimento di Fisica e Astronomia ''G. Galilei'', Universita' di Padova,and INFN, Sezione di Padova, Via F. Marzolo 8, I–35131 Padova, Italy  \label{laveder}}
\date{\today}

\abstract{There are hints, both from cosmology and from neutrino oscillation experiments, that one or two sterile neutrinos 
with eV masses are favored. We consider the implications of combining data from short baseline neutrino experiments with  
cosmological observations. In particular we consider cosmological models where the equation of state of dark energy and the curvature are allowed to vary. We find that the consistency of results from short baseline experiments and cosmology is dependent on the cosmological data sets that are used. When the two are consistent, we find that the inclusion of information from short baseline experiments only mildly affects the preferred values of the cosmological model parameters. A universe with a cosmological constant type dark energy and zero curvature is still fully consistent with the data we use. We also report that a thermalized sterile neutrino in the $\sim$ 10 eV range is fully consistent with a wide range of cosmological data as long as information on the matter power spectrum in not taken into account. }

\keywords{cosmological parameters - dark matter - dark energy - elementary particles - neutrinos}

\titlerunning{Cosmology with sterile neutrinos} 
\authorrunning{J. R. Kristiansen et al.}

\maketitle

\section{Introduction}   

The many links between the very small and the very large are some of the most fascinating aspects of cosmology. To understand the Universe and its history, we must know the fundamental building blocks of matter and their properties. Both particle physicists and cosmologists benefit from the connections between their subjects. For example, cosmological observations  have for a number of years provided the most stringent upper bounds on the neutrino mass scale, and may one day be in a position to determine it \citep{lesgourgues:2012}.  
One could also look at the situation in the opposite way and say that a better knowledge and understanding of the neutrino sector would help us to improve constraints on cosmological parameters. 

The number of neutrino species whose masses are below the GeV scale and
which couple to the Z$^{0}$ boson, i.e., interact weakly, was
determined to be $2.984\pm 008$ from LEP data \citep{nakamura:2010}.  
If there are more
neutrino types than the three we already know about, they must be very
heavy or only couple to gravity, or both.  Neutrinos that do not
participate in the weak interaction are known as sterile.  They
appear in the so-called seesaw mechanism \citep{zuber:2004} for generating low
neutrino masses, and are typically very heavy there, much heavier than
the electroweak scale, in order to explain how low the masses of  the ordinary, active neutrinos are.  
However, as long as it only interacts gravitationally there are no a
priori constraints on the mass of a putative sterile neutrino.

The question of the existence of sterile neutrinos with masses in the eV range has attracted a lot of attention in the last few years 
\citep{donini:2012, giunti:2011a,giunti:2011c, kopp:2011, kopp:2013, giunti:2011b}. There are several experiments hinting at their existence 
\citep{aguilar:2001, aguilar-arevalo:2010, mention:2011, abdurashitov:2006}. 
However, the situation is far from clear, since other experiments disfavour light sterile neutrinos, e.g. \citet{Armbruster:2002mp, adamson:2010}. 
 
On the other hand there have been indications from cosmological probes that a value for the effective number of relativistic particles higher 
than the standard value of $N_{\rm eff} = 3.046$ \citep{mangano:2005, signe:2013} is preferred, the so-called Dark Radiation \citep{giusarma:2012, hamann:2010}.  
Although there are other explanations, one or two eV-mass sterile neutrinos could be the cause of this result.

A number of papers have considered the issue of adding one or two light sterile neutrinos to the cosmological matter budget, 
e.g., \citet{archidiacono:2012, 1302.6720, hamann:2011, hamann:2010}. Most of these studies, however, have added sterile neutrinos to the $\Lambda$CDM model, where the 
cosmological constant is the dark energy. The nature of the dark energy is unknown and the subject of large observational 
projects like the future Euclid mission \citep{amendola:2012}. In this paper we will therefore allow the equation 
of state of the dark energy to be a parameter to be constrained simultaneously with other cosmological parameters and neutrino 
properties. 

In a previous paper \citep{kristiansen:2011} two of us looked into the consequences for cosmology 
if the reactor data really demand the existence of one or two eV-mass sterile neutrinos. The aim of this paper is to improve 
upon this analysis by including the full likelihood function from short base line (SBL) neutrino experiments.

The structure of this paper is as follows. In section 2 we summarize the theoretical background and describe our 
method of analysis. The cosmological data are treated briefly since our method of analysis is standard. Section 3 contains 
our results, and we summarize and conclude in section 4. 

\section{Cosmological models with sterile neutrinos}

Sterile neutrinos with keV masses have been of great interest as dark
matter candidates in cosmology \citep{kusenko:2009} .  Recently it has been suggested
that one or two sterile neutrinos with masses of a few eV lie behind
some puzzling features in neutrino oscillation experiments, see \citet{abazaijan:2012}  for a review.

An important question when we turn to the cosmological implications of
these two scenarios is whether these light, sterile neutrinos were
thermalized in the early universe.  We will assume that they were,
since several studies \citep{hamann:2010,melchiorri:2009,kainulainen:1990}  suggest that this was the case for the masses and mixing parameters we consider, provided that the initial lepton asymmetry was small \citep{hannestad:2012}.   
This means that standard relation
between the sum of the neutrino masses and their contribution to the
cosmic mass density parameter applies. 

Adding light sterile neutrinos  may cause some problems with BBN  
since the increased relativistic energy density results in a larger 
neutron-to-proton ratio, and leads to an increased He$/$H  mass 
fraction.
 Recent analyses \citep{mangano:2011,pettini:2012} found that BBN
constrains the number of relativistic degrees of freedom to be $N_{\rm
eff} < 4$ at 95 \% confidence.  So this means that the 3 active+1 sterile model we consider in this paper is just within the
bounds.

\section{Neutrino oscillation constraints}

We considered the data of SBL neutrino oscillation experiments
in the framework of 3+1 neutrino mixing,
assuming that the mass $m_{4}$ of the fourth neutrino
is much larger than those of the three ordinary neutrinos
which generate the squared-mass differences responsible
for solar and atmospheric neutrino oscillations
(see \citet{GonzalezGarcia:2007ib}).
In this case,
SBL oscillations depend only on the squared-mass difference
$\Delta{m}^2_{41} \simeq m_{4}^2$.
We fitted the
$\nu_{\mu}\to\nu_{e}$ and $\bar\nu_{\mu}\to\bar\nu_{e}$
appearance data of the
LSND \citep{Aguilar:2001ty},
KARMEN \citep{Armbruster:2002mp},
NOMAD \citep{Astier:2003gs},
MiniBooNE \citep{AguilarArevalo:2012va} and
ICARUS \citep{Antonello:2012pq}
experiments,
taking into account the results of
$\nu_{e}$ and $\bar\nu_{e}$
disappearance data \citep{Giunti:2012tn}
(including the
reactor
\citep{mention:2011}
and
Gallium
\citep{Giunti:2010zu}
anomalies)
and the
constraints on
$\nu_{\mu}$ and $\bar\nu_{\mu}$
disappearance
\citep{1302.6720}.
We calculated the Bayesian posterior probability of $\Delta{m}^2_{41}$
through the marginalization over the mixing angles.
This probability is used as a SBL prior in the MCMC analysis of cosmological data.

\begin{figure*}[htb]
\center
\includegraphics[width=0.8\columnwidth]{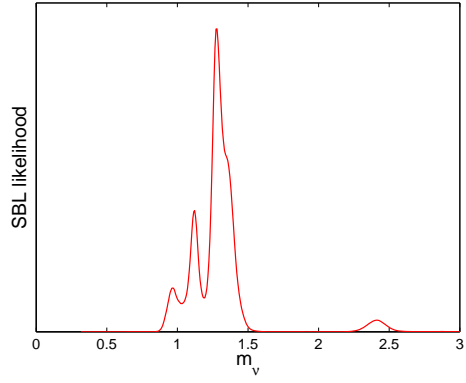}
\caption{Constraints on the mass of a single sterile neutrino from reactor SBL experiments. We use this likelihood distribution as a prior in our cosmological models. As we assume massless active neutrinos, the sterile neutrino mass, \mnu, shown here, is simply taken as $\sqrt{\Delta m_{41}^2}$, where $\Delta m_{41}^2$ is the squared neutrino mass difference measured in the SBL experiments.}
\label{fig:chisq}
\end{figure*}

The posterior from the SBL analysis is shown in Figure \ref{fig:chisq}. The SBL oscillation experiments are sensitive to the squared mass difference between two neutrino mass eigenstates, $\Delta m_{14}^2$. Since we assume the three active neutrino species to be massless, we can use the simple relation $m_\nu = \sqrt{\Delta m_{41}^2}$. 

\section{Cosmological parameter estimation}

For the cosmological parameter analysis we use a modified version of the publicly available Markov chain Monte Carlo code Fortran 90 code CosmoMC \citep{lewis:2002}. We have studied two different basic cosmological models, one minimal, flat \LCDM model, and one model that is extended to allow for a non-zero curvature and dark energy equation of state, $w$, different from -1. The latter model will be denoted \wcdm. We have assumed the three active neutrino species to be massless. Thus, in the \LCDM{} case we have varied the parameters \{$\omega_b$, $\omega_c$, $\theta$, $\tau$, $n_s$, $\ln 10^{10} A_S$\}, and for the \wcdm{} case we have added $\Omega_k$ and $w$ as free parameters. We refer to  \cite{lewis:2002} for exact parameter definitions. 

Then we have added one extra, massive, sterile neutrino species, \snu, to the models, and included the mass of this sterile neutrino, \mnu{} (measured in eV/$c^2$), as a free parameter. For the sterile neutrino mass we analyze two cases. One where we use a flat prior on \mnu{}, and let only the cosmological data constrain its allowed mass range, and one where we impose a prior on \mnu{} from the analysis of short baseline reactor neutrino oscillation experiments. 

To include the SBL results in the cosmological analysis we simply add the log likelihood values from SBL and cosmological data for each step in the MCMC analysis.

The backbone of the cosmological data we use are the9 year CMB data from the WMAP satellite \citep{bennett:2012, hinshaw:2012}, and the likelihood code provided with the data release. In addition we use supernova type 1a data (SN1a) from SDSS \citep{kessler:2009} and matter power spectrum (P(k)) data from the SDSS DR7 LRG sample \citep{reid:2009} and baryonic acoustic oscillation (BAO) data from WiggleZ \citep{blake:2011} and SDSS \citep{percival:2010}. In addition we impose a prior on the Hubble parameter (HST) from \citet{riess:2011} and on the baryon content from nucleosynthesis (BBN) ($\Omega_b h^2 = 0.022 \pm 0.002$). 

We have performed the analysis with the following combinations of data sets: 
\begin{itemize}
\item WMAP9: Only WMAP9 data.
\item Extended: WMAP9 + SN1a + BAO + HST + BBN
\item Extended+Pk: As Extended, but with galaxy power spectrum $P(k)$ data from SDSS instead of the SDSS BAO data. 
\end{itemize}

\section{Results}

\begin{figure*}[htb]
\center
\includegraphics[width=1.5\columnwidth]{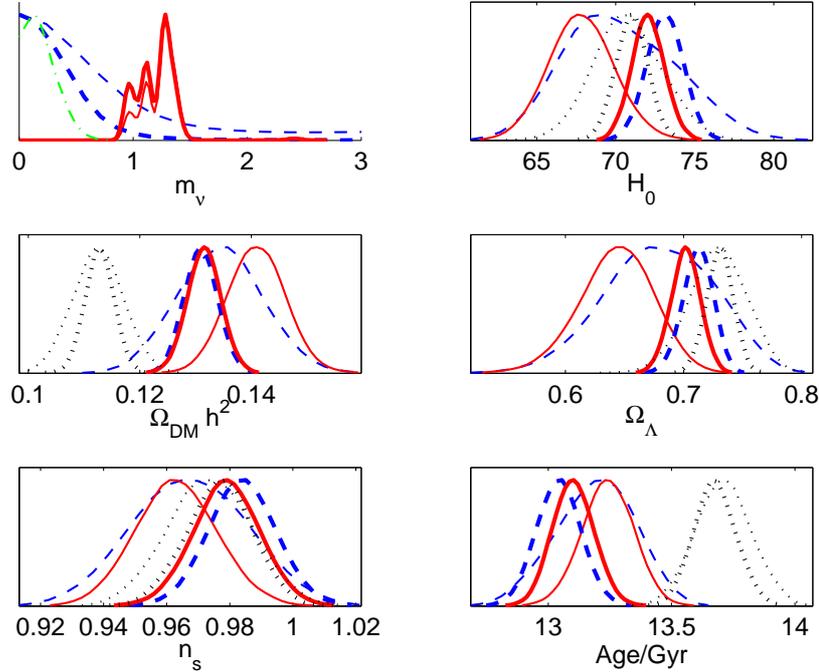}
\caption{Marginalized probability distributions for cosmological parameters in a flat \LCDM+1$\nu_s$ cosmology. Blue, dashed lines: No SBL-prior on \mnu. Red, solid lines: Including a SBL prior on \mnu. Thin lines: WMAP9 data. Thick lines: Extended data. In the \mnu{} panel the green dash-dotted line show the limit on \mnu{} for Extended+Pk data. In the other panels the black, dotted lines show distributions for a flat \LCDM{} model {\it without} any sterile neutrinos, for comparison. }
\label{fig:1D}
\end{figure*}

In Figure \ref{fig:1D} we show 1D marginalized parameter probability distributions obtained for the \LCDM+1\snu{} model for both WMAP9 and Extended data, with and without the SBL prior on \mnu. 

For the sterile neutrino mass we notice that the distribution peaks at zero mass when we only take cosmological data into account. For WMAP9 data only the probability distribution of \mnu{} has a long tail of low but non-zero probability, leaving \mnu{} basically unconstrained. We will discuss this tail more thoroughly later. For Extended data the 95\% region for \mnu{} reaches up to 0.85 eV. When including the SBL prior the \mnu{} distribution peaks at around 1 eV, as this region gives a reasonable fit both to the SBL and cosmological data sets. We see that the SBL prior is consistent with the cosmological analysis using both WMAP9 and Extended data. However, when we include matter power spectrum data and use the Extended+Pk data, the cosmological limits on \mnu{} are tightened to \mnu $<$  0.46 eV (95\% C.L.), which we consider to be incompatible with the SBL prior. We conclude that the strong sterile neutrino mass constraints from the inclusion of $P(k)$ data and the inferred limits from the SBL analysis cannot be valid at the same time. Therefore we do not present results with both Extended+Pk data and the SBL prior.  

Regarding the other cosmological parameters we notice that the inclusion of a sterile neutrino shifts several parameter values significantly compared to a standard \LCDM{} universe without massive neutrinos. For example, the Hubble parameter, $H_0$, and dark energy and dark matter contents are altered, which leads to a shift in the age of the universe to lower values when the sterile neutrino is included. We see only minor shifts in parameter ranges when the adding SBL prior on \mnu is added. The SBL prior does however have a small effect on the values of, for example, $H_0$ and $\Omega_\Lambda$, which are drawn toward smaller values, and, as a result, the age of the universe is shifted towards higher values with the SBL prior included. Using the Extended data set and the SBL prior for the \LCDM+\snu{} universe, we end up with an age estimate of the universe given by 13.10 $\pm$ 0.08 Gyr, which can be compared to 13.68 $\pm$ 0.08 Gyr when using the same cosmological  data sets for a \LCDM{} model without sterile neutrinos.

\begin{figure*}[htb]
\center
\includegraphics[width=1.5\columnwidth]{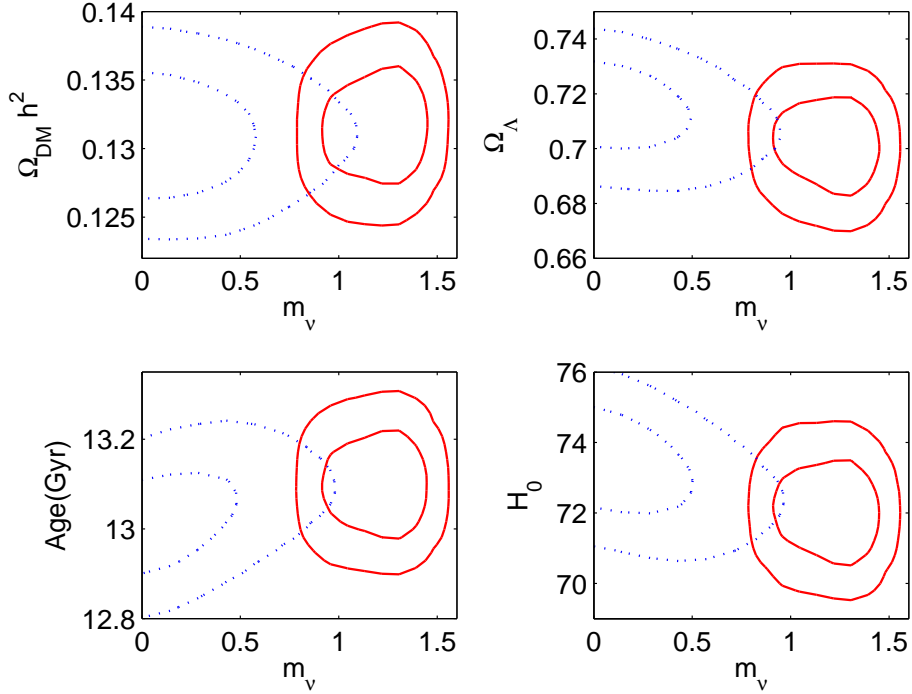}
\caption{2D probability contours for selected parameters in the case of a flat \LCDM+1\snu{} model using Extended data. The contours indicate 95\% and 68\% confidence intervals. Blue lines show results without any SBL prior on \mnu, while the red contours indicate the results when including the SBL prior in the analysis. }
\label{fig:2D}
\end{figure*}

The reason for these shifts can be seen in Figure \ref{fig:2D}. Here we show two-dimensional confidence intervals (95\% and 68\%) for a few parameter combinations when using the flat \LCDM+1\snu{} model and Extended data. We see that there are degeneracies between \mnu and $\Omega_\Lambda$ and $H_0$, and as a consequence also the age of the universe. 

\begin{figure*}[htb]
\center
\includegraphics[width=1.5\columnwidth]{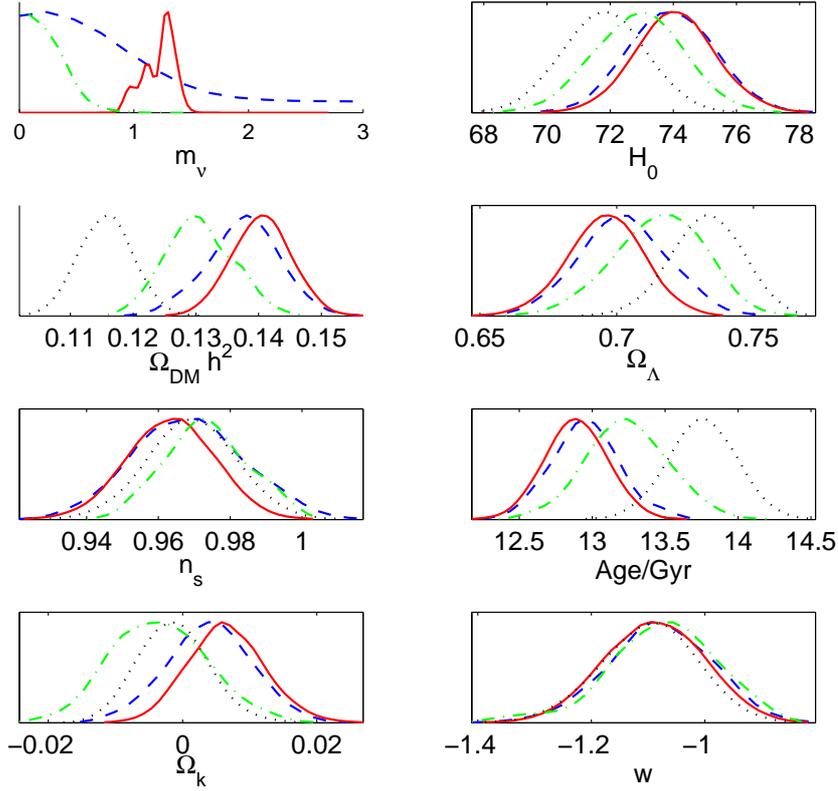}
\caption{Marginalized probability distributions for cosmological parameters in a \wcdm+\snu{} cosmological model. Line colors corresponding to Figure \ref{fig:1D}. As this extended model is poorly constrained by WMAP9 data alone, we only show results from the Extended analysis. For the Extended+Pk cosmological data (green dash-dotted lines) we only show results without a SBL prior. }
\label{fig:w_1D}
\end{figure*}

Next, we turn to the extended cosmological model, \wcdm{}, where we allow the equation of state of dark energy, $w$, and the curvature of the universe, $\Omega_k$ to vary. In Figure \ref{fig:w_1D} we show the probability distributions for the different parameters in this case. As these models are very poorly constrained by WMAP9 data only, we only show results from the Extended and Extended+Pk data sets. 

For the distribution of \mnu{} analyzed with the Extended data set we see that the distribution falls off in an almost Gaussian manner up to \mnu $\sim$ 2 eV, and that it retains a small, non-zero probability for higher masses. We will discuss this high-mass tail further later  on. The SBL prior is fully consistent with the extended data set, and adding this prior, constrains the neutrino mass to be between 0.96 eV and 1.40 eV at 95\% C.L. Again, including matter power spectrum data, using the Extended+Pk data set we get limits on \mnu{} that are incompatible with the SBL prior (\mnu < 0.55  eV (95\% C.L.)). Therefore we do not show results where the SBL prior is added to the Extended+Pk analysis. 

For the other parameters we find shifts of the order of 1$\sigma$ by adding a massive sterile neutrino, compared to a model without sterile neutrinos. Adding a sterile neutrino the preferred model is drawn towards less dark energy density, and more dark matter. The distributions of $\Omega_k$ and $w$ shift by less than 1$\sigma$ for all the combinations of models and data sets shown here. This may be surprising, as we have found in a previous work that a sterile neutrino in the 1 eV range gives a preference for $w < -1$. In that analysis, however, we did not make use of BAO data. The inclusion of BAO data puts an additional strong handle on $w$, which makes it strongly constrained, regardless of the inclusion of additional massive neutrinos.  

Adding the SBL prior on the neutrino mass does not affect the distributions significantly. We do however see small shifts in parameters like the scalar spectral index, $n_s$, where the significance of $n_s<1$ increases from 2.1$\sigma$ to 3.0$\sigma$ with the inclusion of an SBL prior. This is important for inflationary models of the early universe. 

In the \wcdm+\snu{} model we infer an age of the universe of 12.89 $\pm$ 0.22 Gyr when using the extended data set and the SBL prior. 

\begin{figure*}[htb]
\center
\includegraphics[width=1.5\columnwidth]{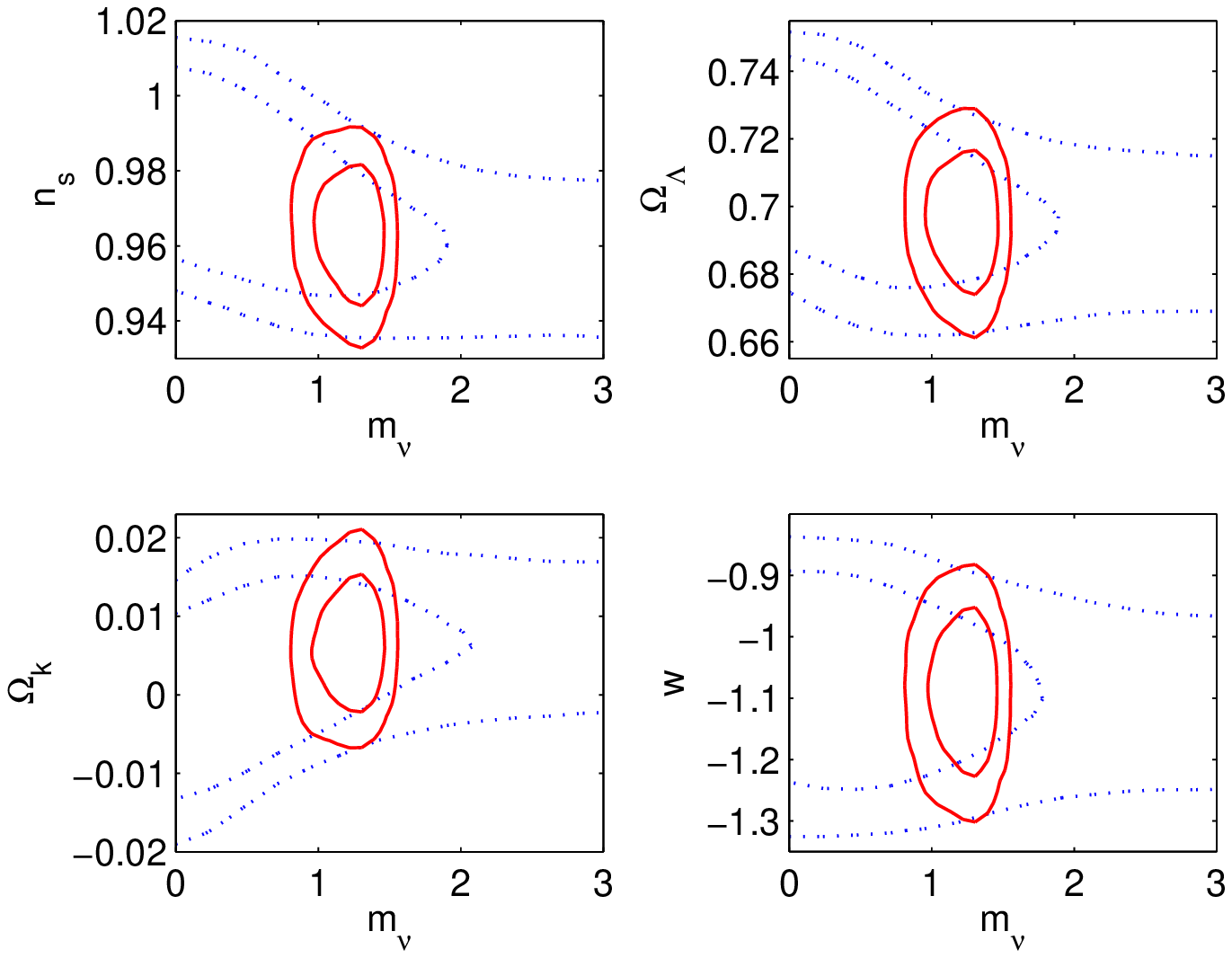}
\caption{2D 95\% and 68\% probability contours for selected parameters in a \wcdm+\snu cosmological model, using Extended data. Blue contours: No SBL prior. Red lines: With the SBL prior included. }
\label{fig:w_2D}
\end{figure*}

In Figure \ref{fig:w_2D} we show 2D distributions for a few parameter combinations in the case of a \wcdm{} model analyzed with the Extended data set, with and without the SBL prior. We notice that there is basically no correlation between the sterile neutrino mass and $w$, but that there are correlations between $m_\nu$ and $n_s$, $\Omega_\Lambda$ and $\Omega_k$.  In particular we see that the anti-correlation between $n_s$ and \mnu{} show why SBL prior results in a stronger preference for $n_s < 1$.

\begin{figure*}[htb]
\center
\includegraphics[width=0.9\columnwidth]{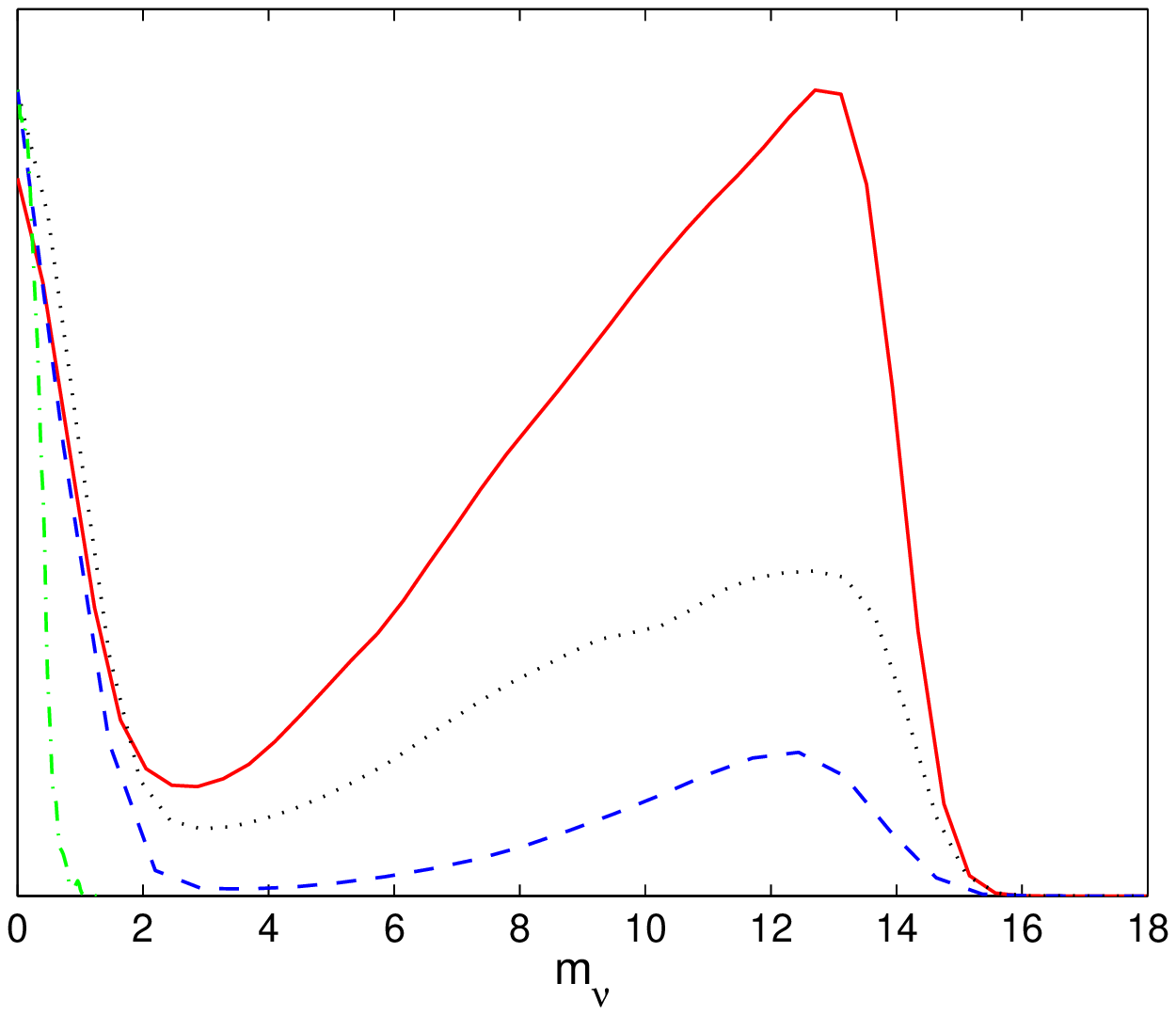}
\caption{Probability distributions for \mnu{} for different combinations  of data sets and cosmological models without SBL prior. Red, solid line: \LCDM+\snu{} model, WMAP9 data.  Black, dotted line: \LCDM+\snu{} model, Extended data. Blue, dashed line: \wcdm{} model, Extended data. Green, dash-dotted line: \wcdm{} model, Extended+Pk data.}
\label{fig:sm}
\end{figure*}

Again, we notice the tail of an allowed region towards \mnu $> 2$ eV for the analysis using the Extended data set. In Figure \ref{fig:w_2D} the 95\% region is indeed extending beyond \mnu = 3 eV. In Figure \ref{fig:sm} we show the marginalized probability distribution of \mnu{} for different combinations of models and data sets, without using the SBL prior. For the Extended+Pk data, the probability for $m_\nu > 3$ eV remains zero. However, for both the WMAP9 and Extended data sets, a new high-probability region appears in the range $\sim$ 4 eV - 15 eV. Indeed, for the  \LCDM+\snu model analyzed with WMAP9 data only, the preferred value of \mnu{} peaks at 13 eV. When the Extended data is used, the significance of this peak is lowered, and with the inclusion of matter power spectrum data the high-mass peak vanishes.  

Does this peak hint towards the existence of a sterile neutrino with a mass in the 10 eV range? Not really. In both the WMAP9 and Extended data sets the only direct handle we have on matter clustering comes from the CMB power spectrum. Neutrino masses affect the CMB power spectrum in several ways \citep{lesgourgues:2006}, but one of the most important effects stems from the shift of the time of matter-radiation equality ($t_{eq}$). Neutrinos with masses less than $\sim 2$ eV will still be relativistic at this time. Therefore, increasing the fraction of dark matter in the form of neutrinos, $f_\nu$, will shift $t_{eq}$ to later times, which in turn will affect the structure formation. However, neutrinos with masses $\gtrsim 2$ eV will already be not-relativistic at $t_{eq}$, and will therefore {\it not} shift $t_{eq}$ relative to a pure cold dark matter model. In this regard, neutrinos in the $\sim$ 10 eV range, corresponding to the high-mass peaks in Figure \ref{fig:sm}, will therefore act similar to a cold dark matter component when it comes to the effect of $t_{eq}$ on the CMB. When \mnu $\sim$ 10 eV, $f_\nu$ will approach 1 for fully thermalized sterile neutrinos (the exact value of course depending on the total dark matter content), meaning that almost all of the dark matter is in the form of sterile neutrinos. Then the normal upper bounds on the dark matter content of the universe will apply, explaining why the \mnu{} probability distributions falls of to zero for \mnu $\sim$ 15 eV. 

When we include the matter power spectrum, using the Extended+Pk data, the situation changes. Here the large thermal velocities of the neutrinos give the neutrinos a free-streaming scale that suppresses structure growth at small scales. The signature of a universe with $f_\nu \sim 1$ will therefore look dramatically different from a universe with $f_\nu \sim 0$.  

\begin{figure*}[htb]
\center
\includegraphics[width=1.9\columnwidth]{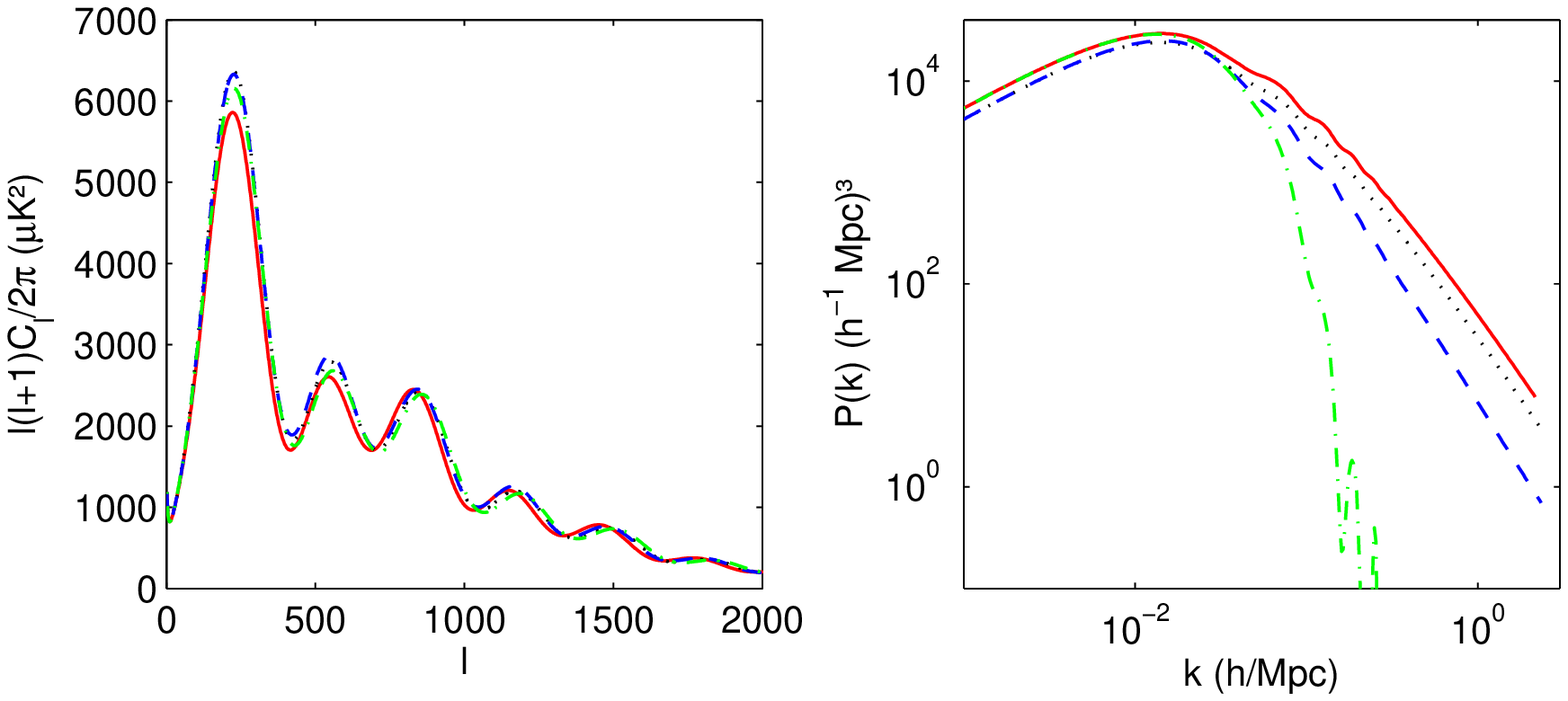}
\caption{CMB and matter power spectrum for \LCDM{} and \LCDM+\snu{} models with different values of \mnu. The total dark matter density is kept constant, so the plots show the effect of changing different amounts of cold dark matter into hot dark matter in form of fully thermalized sterile neutrinos. The other cosmological parameters are kept to standard, concordance values. Red, solid line: \LCDM{} model.  Black, dotted line: \LCDM+\snu{} model with \mnu=1 eV. Blue, dashed line: \LCDM+\snu{} model with \mnu=3 eV. Green, dash-dotted line: \LCDM+\snu model with \mnu=10 eV. }
\label{fig:pk}
\end{figure*}

In Figure \ref{fig:pk} we show the CMB and matter power spectra for cosmological models with different masses of the sterile neutrinos. The other cosmological parameters are set to concordance values. The model with \mnu = 10 eV model therefore corresponds to a universe with a $f_\nu$ close to 1. For the CMB power spectrum, the model with \mnu = 10 eV actually has a larger resemblance to the standard \LCDM model than the models with \mnu = 1 eV or \mnu = 3 eV. This can be understood by the reasoning above, regarding the neutrino mass effect on $t_{eq}$. However, when we turn to the matter power spectrum, we see that the \mnu = 10 eV model gives an extreme suppression of small scale structure, in total disagreement with all matter power spectrum data. Also, such a high value of \mnu{} is in disagreement with the SBL results used here. 

Thus, we regard the high-mass peaks of \mnu{} not to be compatible with our observed universe, at least not within our model framework. In the analysis of the data from WMAP9 and the Extended data sets shown above, we have therefore imposed a cutoff at $m_\nu = $ 5 eV in the analysis of the Monte Carlo chains, to avoid the inclusion of the region of parameter space with an artificially high \mnu. 

\section{Discussion and conclusions}

Our analyses show consistency between cosmology and SBL data when the cosmological data consists of the combination WMAP9-SN1a+BAO+HST+BBN. The small peak in the SBL prior seen around 2.5 eV is removed by cosmology, but the higher peak around 1 eV remains. 
However, cosmology becomes inconsistent with SBL when the BAO data are replaced with the galaxy power spectrum estimated from the SDSS LRG sample. The latter result depends on the relation between the distributions of galaxies and the dark matter, a relation which is non-trivial and not understood in detail. Whether a better understanding of the small-scale clustering of galaxies and matter will leave room for a sterile neutrino with mass $\sim 1$ eV remains to be seen. There is, however, no reason to doubt that the galaxy power spectrum data exclude the peak around 10 eV in the probability distribution for $m_\nu$ that appears when using CMB + background evolution data only. 

While \cite{kristiansen:2011} found a preference for the equation of state of dark energy $w<-1$ 
when a sterile neutrino was included, in this work we find that the cosmological constant lies within 
1$\sigma$ contour. The BAO data play a crucial role for this result by effectively removing the degeneracy between $m_\nu$ and $w$.

Our results are consistent with those of \citep{1302.6720}. They considered a wider range of 
cases for sterile neutrinos, e.g., both 3+1 and 3+2, and the possibility of incomplete thermalization, 
but did not allow the equation of state of dark energy to vary, and assumed a spatially flat universe. 
We have allowed $w$ and the spatial curvature to vary, but have considered the 3+1 case with 
a fully thermalized sterile neutrino. Both analyses agree that a sterile neutrino with mass 
$\sim 1$ eV is consistent with cosmology.

The CMB on its own has a limited ability to constrain neutrino masses \citep{lesgourgues:2006}, even with 
the precision we can expect from the soon-to-be released Planck data. This is well illustrated by the peak around 12 eV for the sterile neutrino mass we obtained when not using data on the matter power spectrum. But by removing parameter 
degeneracies when combined with other probes, we can expect better constraints on $m_\nu$. It will be interesting to see if cosmology can be consistent with a 1 eV sterile neutrino when Planck data are combined with a cleaner probe of the clustering of matter than the galaxy power spectrum.

\begin{acknowledgements}
Large parts of the calculations in this work were performed on the Abel Cluster, owned by the University of Oslo
and NOTUR. We thank Nicolaas Groeneboom for useful help. 
\end{acknowledgements}

\bibliographystyle{aa} 

\end{document}